\documentclass[useAMS,usenatbib]{mn2e}

\usepackage{epsfig}
\usepackage{amsfonts}
\usepackage[usenames]{color}
\usepackage[usenames,dvipsnames]{xcolor}
\usepackage{url}
\usepackage{subfigure}
\usepackage{times,graphicx,amsmath,amsfonts,amssymb,aas_macros,epstopdf,hyperref}
\usepackage[normalem]{ulem}
\usepackage[T1]{fontenc}
\usepackage{multirow}
\usepackage{aecompl}

\usepackage{hyperref}
\hypersetup{
    colorlinks=true,
    linkcolor=blue,
    citecolor=blue,
}

\def\etal{{\frenchspacing\it et al.}}
\def\ie{{\frenchspacing\it i.e.}}
\def\eg{{\frenchspacing\it e.g.}}

\def\be{\begin{equation}}
\def\ee{\end{equation}}
\def\ba{\begin{eqnarray}}
\def\ea{\end{eqnarray}}

\newcommand{\mpcoh}{\,h^{-1}\,{\rm Mpc}}

\def\d{{\rm d}}

\def\LaTeX{L\kern-.36em\raise.3ex\hbox{a}\kern-.15em
    T\kern-.1667em\lower.7ex\hbox{E}\kern-.125emX}

\begin{document}

\voffset-1.25cm
\title[Tomographic analysis of anisotropic galaxy clustering]{The clustering of galaxies in the completed SDSS-III Baryon Oscillation Spectroscopic Survey: a tomographic analysis of structure growth and expansion rate from anisotropic galaxy clustering}
\author[Wang \etal]{
\parbox{\textwidth}{
Yuting Wang$^{1,2}$\thanks{\url{Email: ytwang@nao.cas.cn}}, Gong-Bo Zhao$^{1,2}$\thanks{\url{Email: gbzhao@nao.cas.cn}}, Chia-Hsun Chuang$^{3}$, Marcos Pellejero-Ibanez$^{4,5}$, Cheng Zhao$^{6,1}$, Francisco-Shu Kitaura$^{3}$, Sergio Rodriguez-Torres$^{7,8}$}
\vspace*{15pt} \\
$^{1}$ National Astronomy Observatories, Chinese Academy of Science, Beijing, 100012, P. R. China \\
$^{2}$ Institute of Cosmology \& Gravitation, University of Portsmouth, Dennis Sciama Building, Portsmouth, PO1 3FX, UK \\
$^{3}$ Leibniz-Institut f\"ur Astrophysik Potsdam (AIP), An der Sternwarte 16, D-14482 Potsdam, Germany \\
$^{4}$ Instituto de Astrof\'{\i}sica de Canarias (IAC), C/V\'{\i}a L\'actea, s/n, E-38200, La Laguna, Tenerife, Spain \\
$^{5}$ Departamento Astrof\'{\i}sica, Universidad de La Laguna (ULL), E-38206 La Laguna, Tenerife, Spain \\
$^{6}$ Tsinghua Center of Astrophysics and Department of Physics, Tsinghua University, Beijing 100084, China \\
$^{7}$ Campus of International Excellence UAM+CSIC, Cantoblanco, E-28049 Madrid, Spain \\
$^{8}$ Departamento de F\'{\i}sica Te\'orica, Universidad Aut\'onoma de Madrid, Cantoblanco, E-28049, Madrid, Spain
}
\date{\today} 
\pagerange{\pageref{firstpage}--\pageref{lastpage}}

\label{firstpage}

\maketitle


\begin{abstract} 

We perform a tomographic analysis of structure growth and expansion rate from the anisotropic galaxy clustering of the combined sample of Baryon Oscillation Spectroscopic Survey (BOSS) Data Release 12, which covers the redshift range of $0.2<z<0.75$. In order to extract the redshift information of anisotropic galaxy clustering, we analyse this data set in nine overlapping redshift slices in configuration space and perform the joint constraints on the parameters $\big\{D_V \times \left(r_d^{\rm fid}/r_d\right), F_{\mathrm{AP}}, f\sigma_8\big\}$ using the correlation function multipoles. The analysis pipeline is validated using the MultiDark-Patchy mock catalogues. We obtain a measurement precision of $1.5\%-2.9\%$ for $D_V \times \left(r_d^{\rm fid}/r_d\right)$, $5.2\%-9\%$ for $F_{\mathrm{AP}}$ and $13.3\%-24\%$ for $f \sigma_8$, depending on the effective redshift of the slices. We report a joint measurement of $\big\{D_V \times \left(r_d^{\rm fid}/r_d\right), F_{\mathrm{AP}}, f\sigma_8\big\}$ with the full covariance matrix in nine redshift slices. We use our joint BAO and RSD measurement combined with external datasets to constrain the gravitational growth index $\gamma$, and find $\gamma=0.656 \pm 0.057$, which is consistent with the $\Lambda$CDM prediction within 95\% CL.
\end{abstract}

\begin{keywords} 

large scale structure of Universe; redshift space distortions; baryon acoustic oscillations

\end{keywords}

\section{Introduction}
\label{sec:intro}

The galaxy redshift survey is a powerful probe for the nature of dark energy (DE) and gravity, both of which are crucial to understanding the accelerating expansion of the Universe at late times, as discovered by observations of type Ia supernovae \citep{Riess, Perlmutter}. Redshift surveys allow us to measure the cosmic expansion history and structure growth simultaneously by statistically analysing the three-dimensional clustering of the galaxies in terms of the correlation function in configuration space or the power spectrum in Fourier space \citep{Cole1995, Peacock2001, Cole2005, Hawkins2003, Eisenstein2005, Okumura2008, Percival2009}.

The observed baryon acoustic oscillations (BAO), as a ``standard ruler'', in the correlation function or power spectrum can be used to probe the cosmic expansion history, since the signal is robust to systematic uncertainties \citep{Eisenstein2004, Padmanabhan2009, Mehta2011, Vargas2016}. The measured BAO scales in the radial and transverse directions from the anisotropic galaxy clustering provide an estimate of the Hubble parameter, $H(z)$, and angular diameter distance, $D_A(z)$, respectively. 

The anisotropy in the galaxy clustering is partially due to the Alcock-Paczynski (AP) effect \citep{Alcock1979}, which arises from assuming a wrong cosmology to convert redshifts to distances for the clustering analysis. The distortion of distances along and perpendicular to the line-of-sight (LOS) direction depends on the offset in the Hubble parameter, $H(z)$ and the angular diameter distance, $D_A(z)$ respectively. Therefore, measuring the relative distortion in the radial and transverse directions provides a probe of $D_A(z)$ and $H(z)$. Another source of anisotropy in galaxy clustering arises from the large-scale redshift-space distortions (RSD) \citep{Kaiser1987}, which is the consequence of peculiar motions of galaxies. Galaxies tend to infall towards the local over-density region, thus the clustering along the LOS is enhanced. The measurement of RSD can provide us with the growth history of large-scale structure, which is parametrized as $f(z)\sigma_8(z)$, here $f(z)$ is the growth rate, and $\sigma_8(z)$ is the linear-theory root mean square (rms) mass fluctuations in spheres of radius 8 $\mpcoh$ \citep{Song2009, Percival2009}, and can be used to distinguish various theoretical models, including tests of gravity \citep{Song2009, Raccanelli2013, Samushia2013, Beutler2014, Mueller2016}. 

The Baryon Oscillation Spectroscopic Survey (BOSS) \citep{Dawson2013}, part of SDSS-III \citep{Eisenstein2011}, has provided the final Data Release 12 (DR12) \citep{Alam}, which is the largest data set for galaxy redshift surveys to date, and includes spectroscopic redshifts of more than a million galaxies. \citet{Hector2016} carried out a RSD analysis in Fourier space using the DR12 CMASS catalogue in the redshift range of $0.43<z<0.75$ and the LOWZ catalogue in the redshift range of $0.15<z<0.43$. Using these samples, \citet{Pellejero2016} and \citet{Chuang2016} performed an analysis of the anisotropic clustering in configuration space. Using the ``combined'' sample of BOSS DR12 covering the redshift range of $0.2<z<0.75$ \citep{Alam2016}, a joint analysis of cosmic expansion rate and growth structure from the anisotropic clustering of galaxy was performed in \citep{Alam2016, Beutler2017, Satpathy2017, Ariel2017I, Ariel2017II, Grieb2017} in three redshift slices of $0.2<z<0.5$, $0.4<z<0.6$ and $0.5 <z<0.75$. In order to extract the lightcone information of galaxy clustering, we performed the BAO analysis by splitting the sample into multiple overlapping redshift slices in configuration space \citep{Wang2017} and in Fourier space \citep{Zhao2017}, respectively. 

In this paper, we perform a joint BAO and RSD analysis in nine overlapping redshift slices using the correlation function multipoles from the pre-reconstructed catalogues of BOSS DR12 and the data covariance matrix estimated from the MultiDark-Patchy (MD-P) mock catalogues \citep{Kitaura2016} (see \citet{Wang2017} for details of the correlation function measurements). We adopt the ``Gaussian streaming model'' (GSM) developed in \citep{Reid2011} as the template. We review GSM and the fitting method in Section \ref{Method}. Our results are presented in Section \ref{result}. Section \ref{conclusion} is devoted to the conclusion. In this paper, we use a fiducial $\Lambda$CDM cosmology with parameters:  $\Omega_m=0.307, \Omega_bh^2=0.022, h=0.6777, n_s=0.96,  \sigma_8=0.8288$. The comoving sound horizon in this cosmology is $r_d^{\rm fid}=147.74 \,\rm Mpc$.

\begin{figure}
\centering
{\includegraphics[scale=0.3]{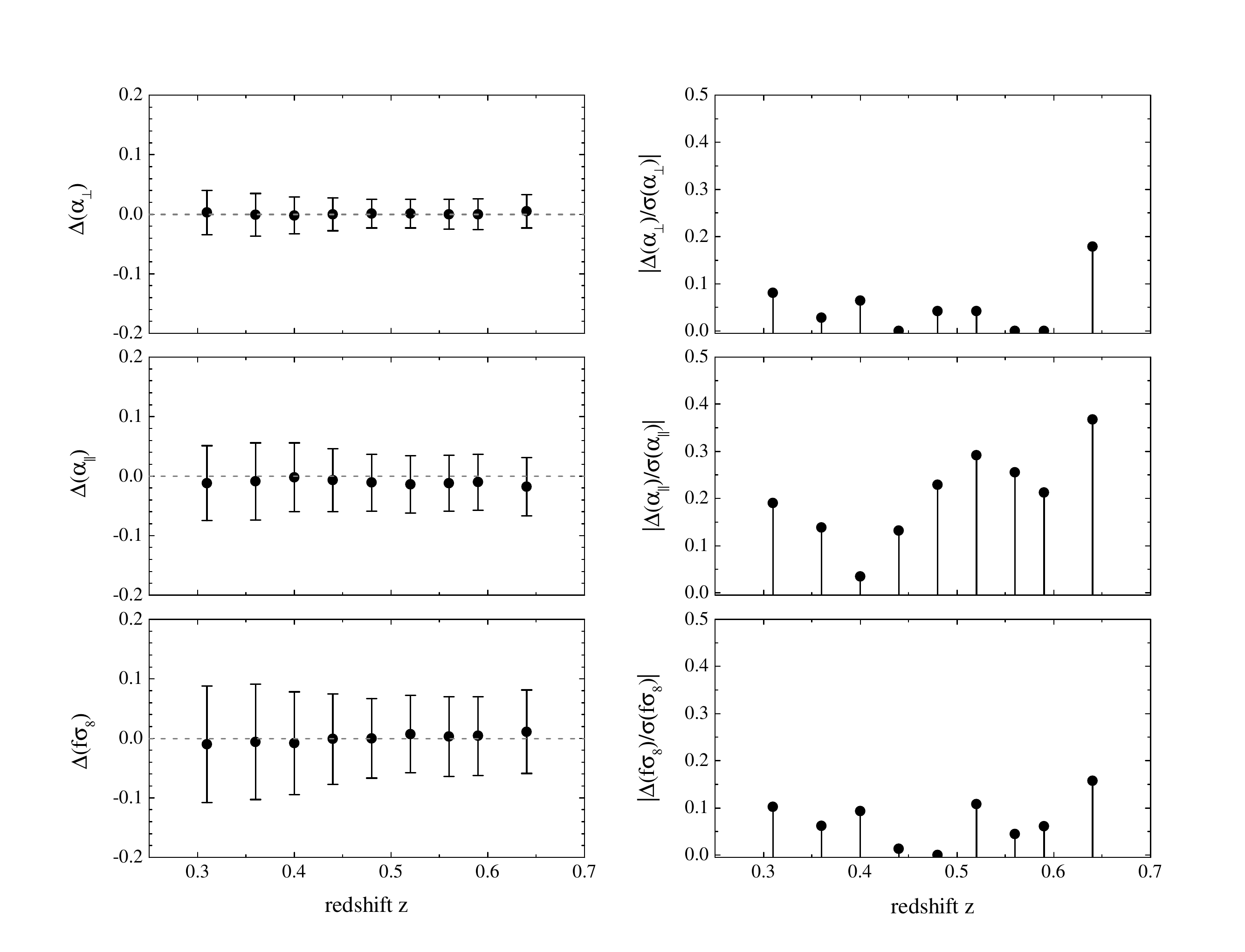}}
\caption{Left panels: The difference between the mean and the expected values. Right panels: the significance of the bias in terms of the 68\% CL uncertainty.}
\label{fig:combias}
\end{figure}

\begin{figure*}
\centering
{\includegraphics[scale=0.22]{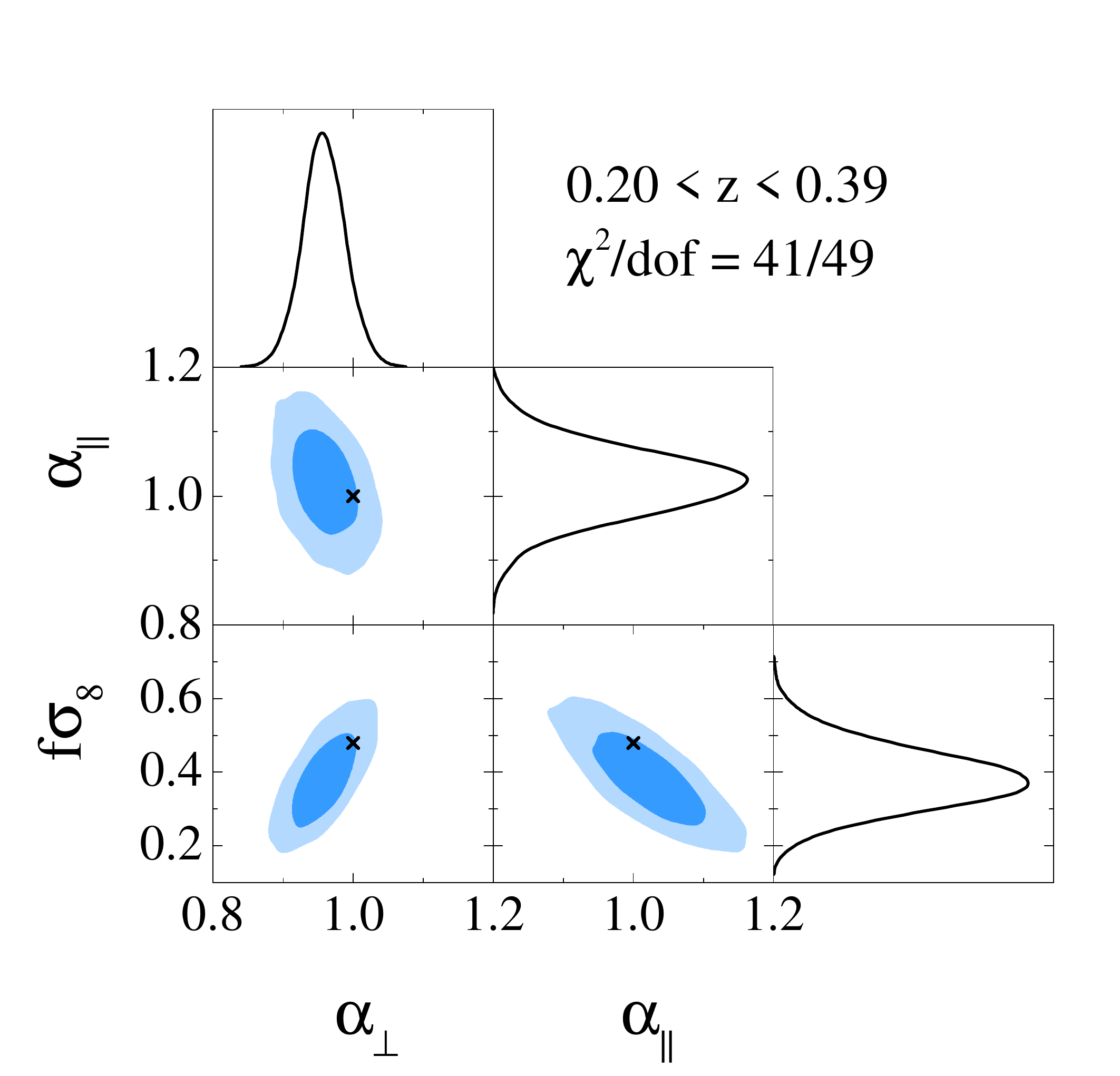}} {\includegraphics[scale=0.22]{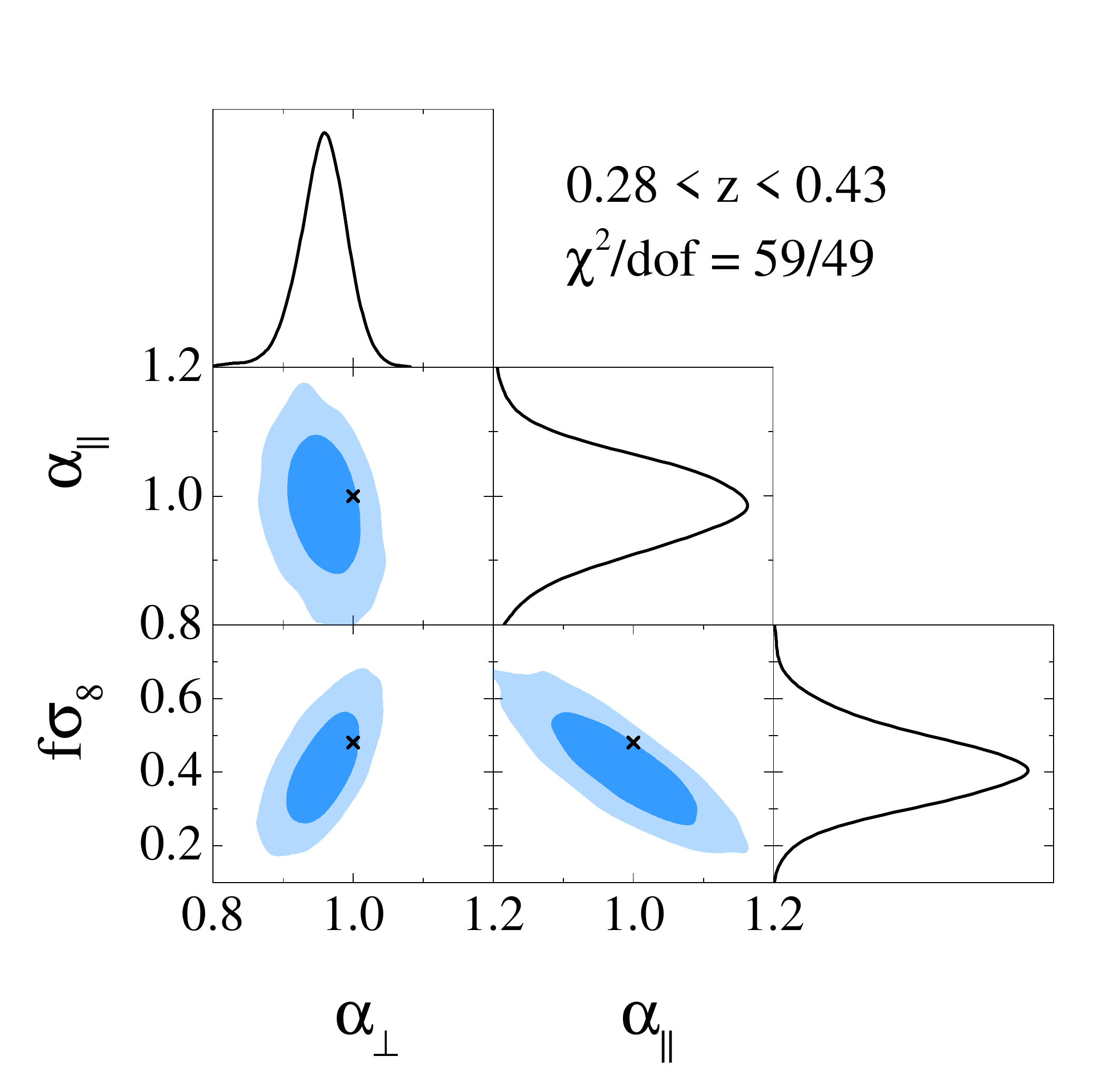}} {\includegraphics[scale=0.22]{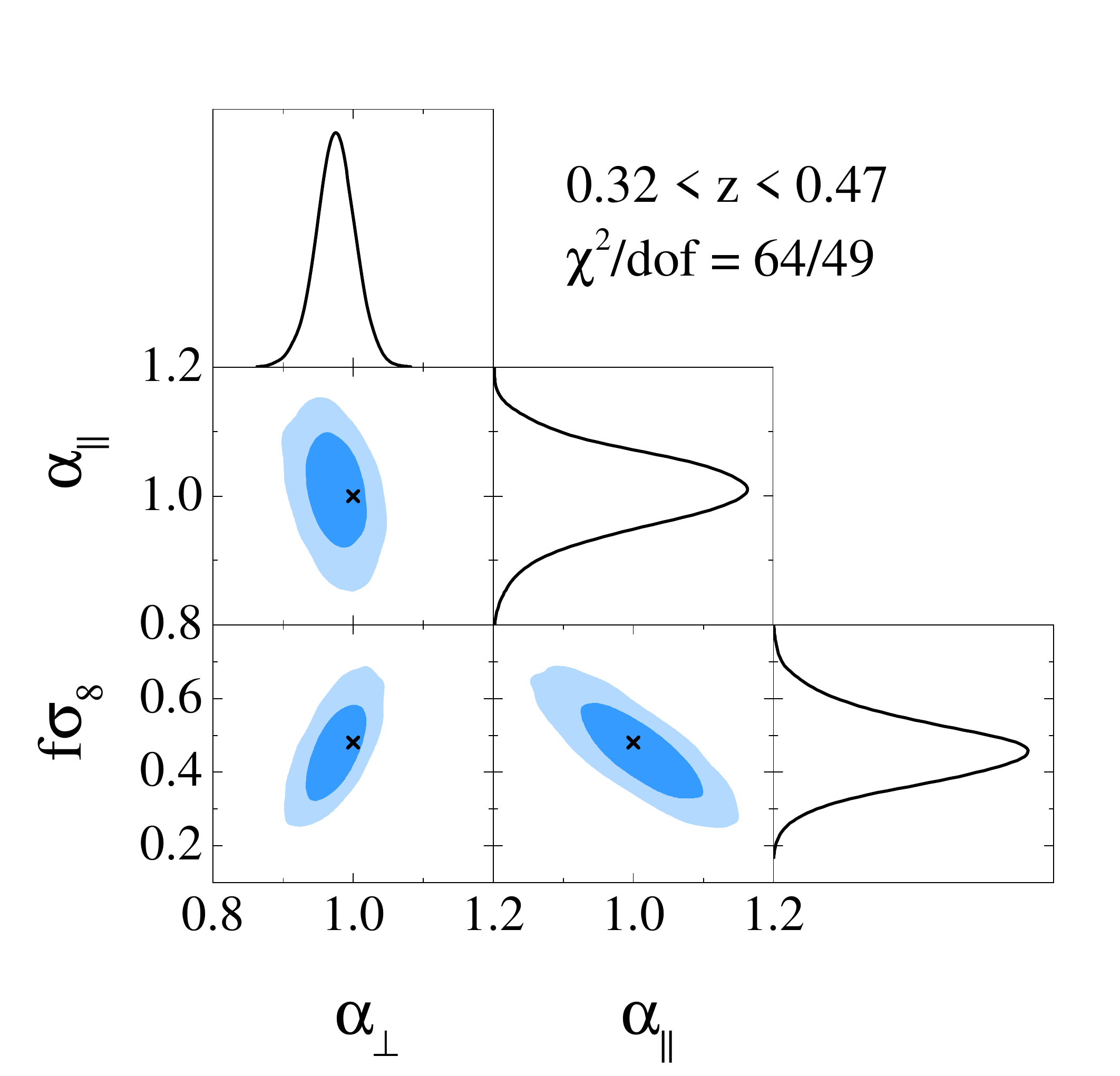}} \\
{\includegraphics[scale=0.22]{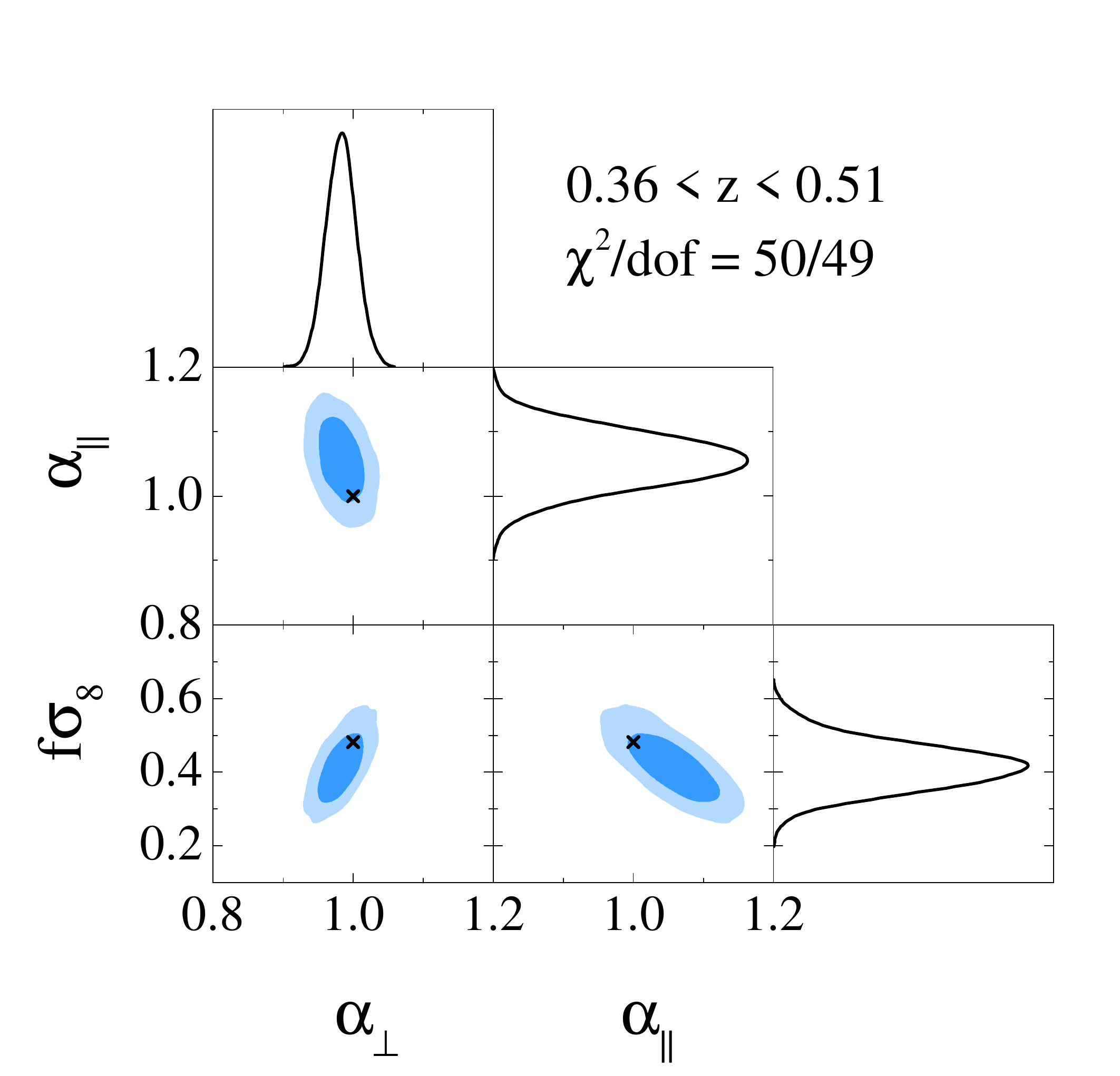}} {\includegraphics[scale=0.22]{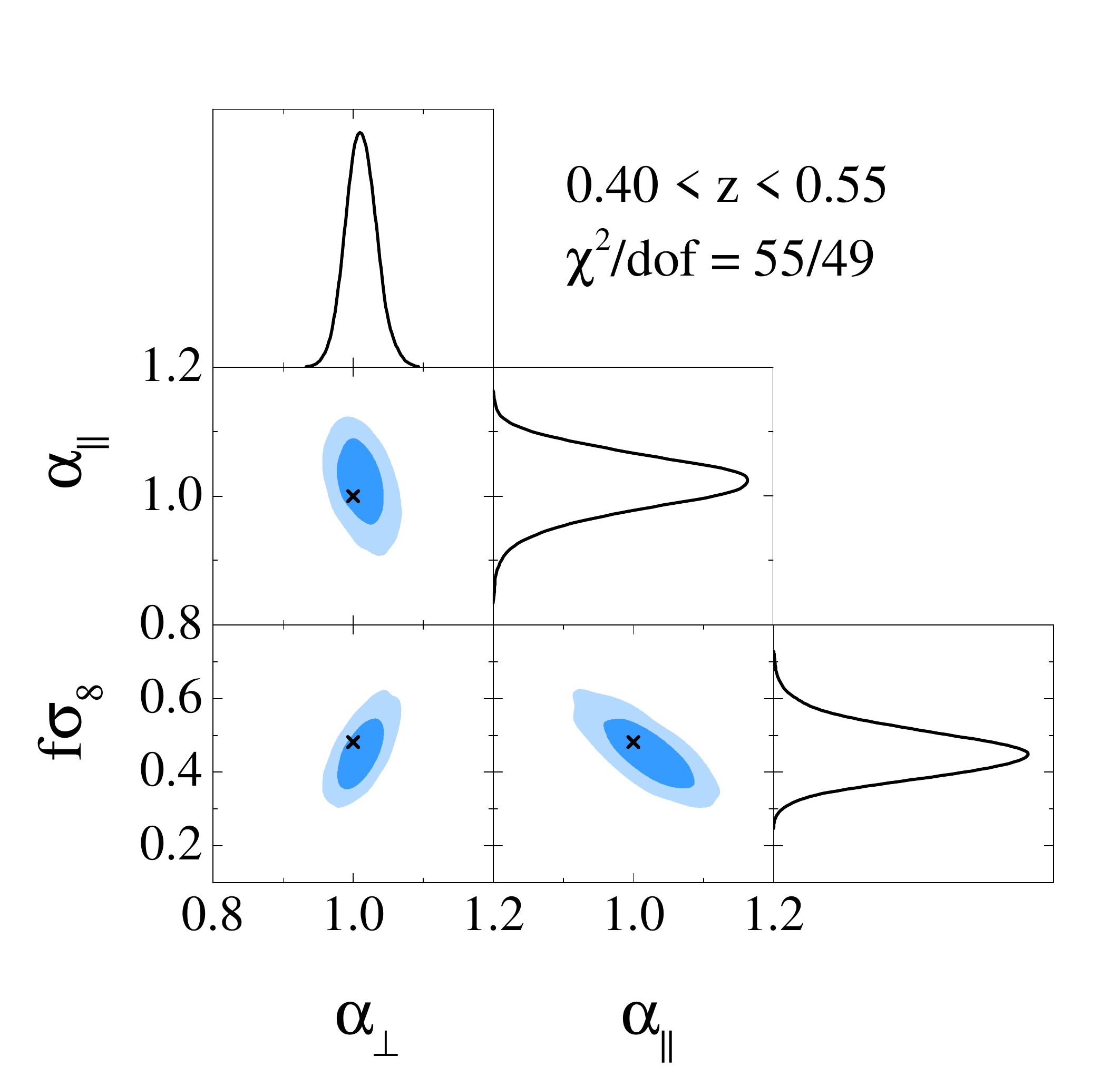}} {\includegraphics[scale=0.22]{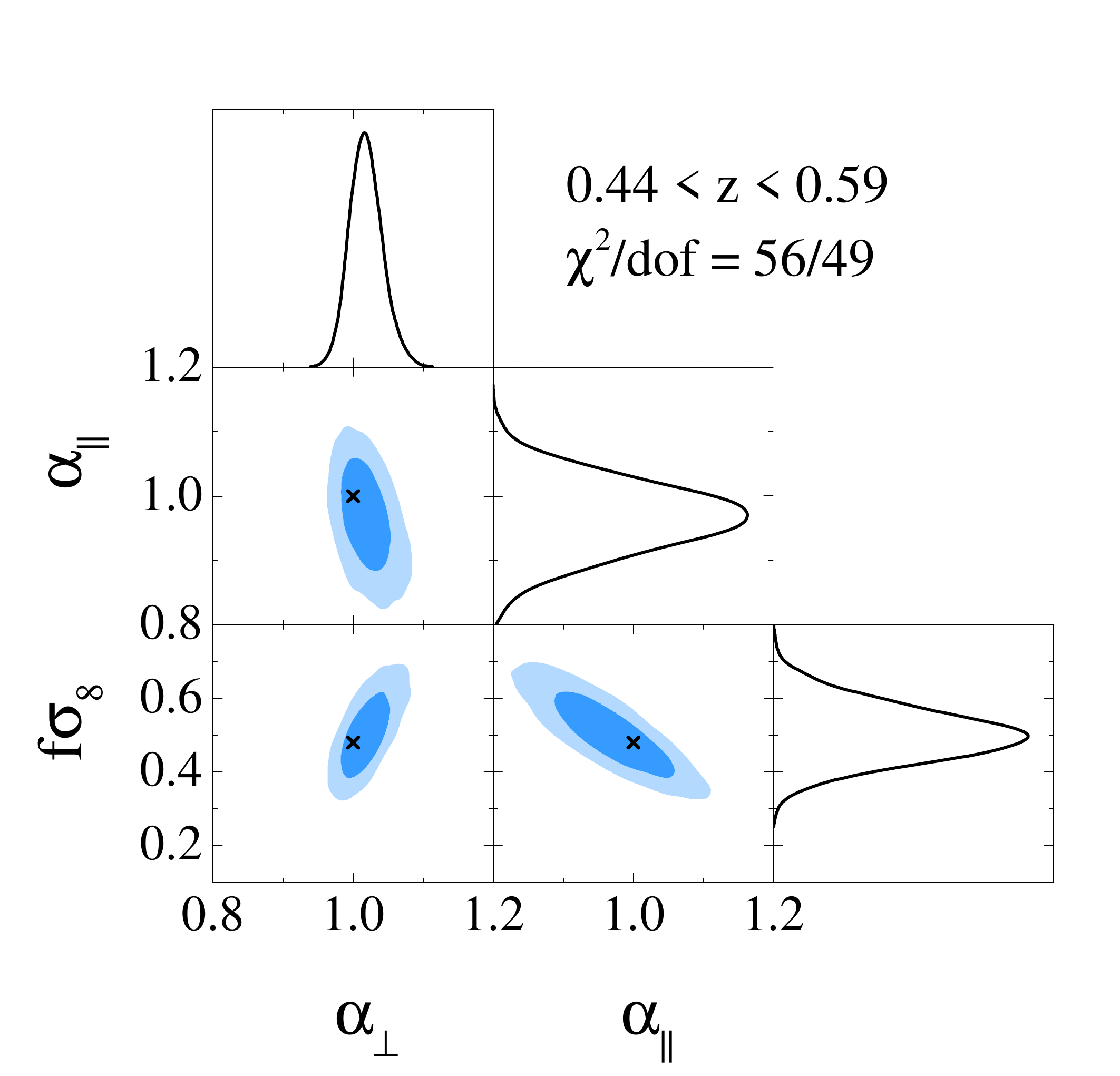}} \\
{\includegraphics[scale=0.22]{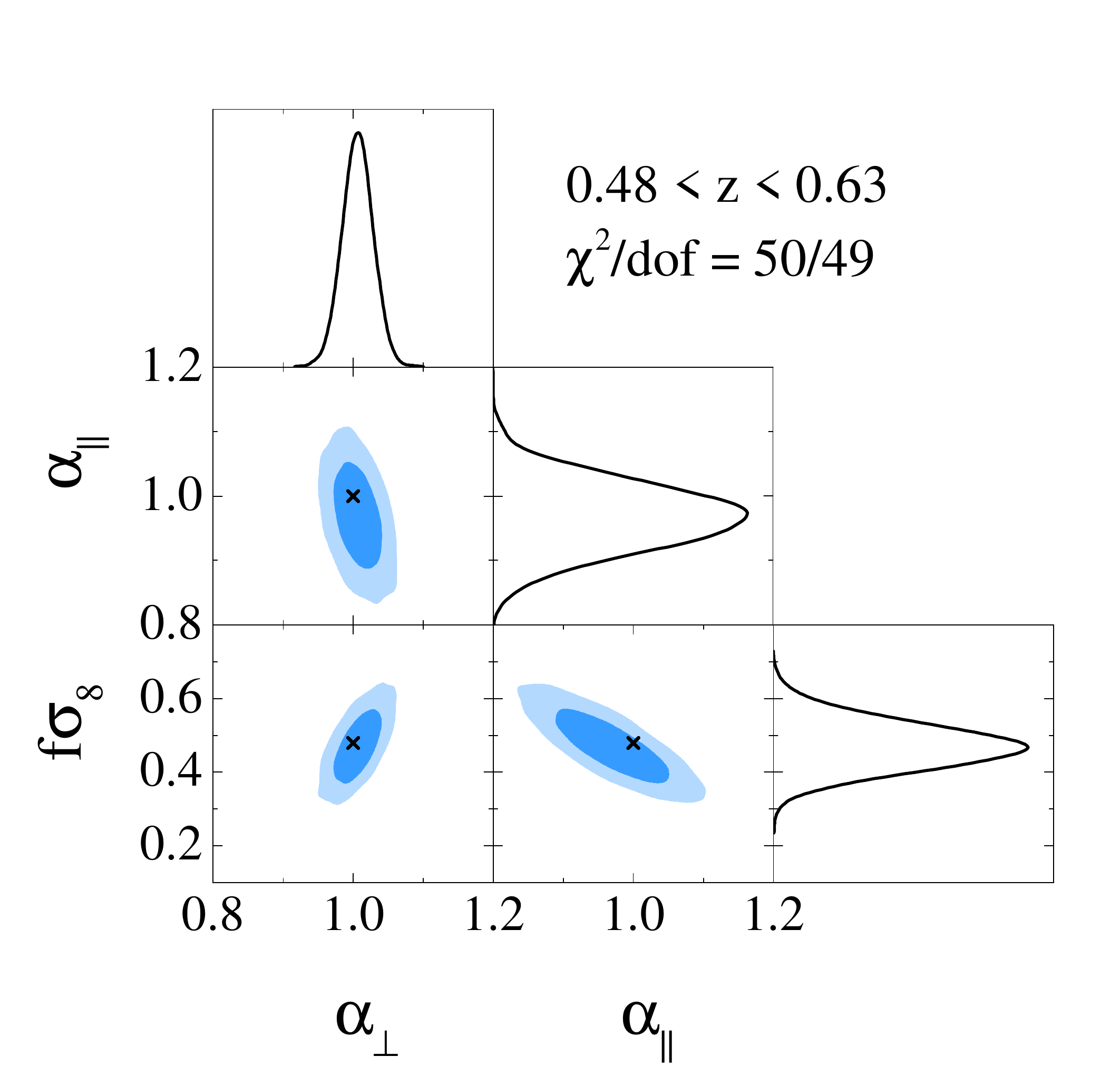}} {\includegraphics[scale=0.22]{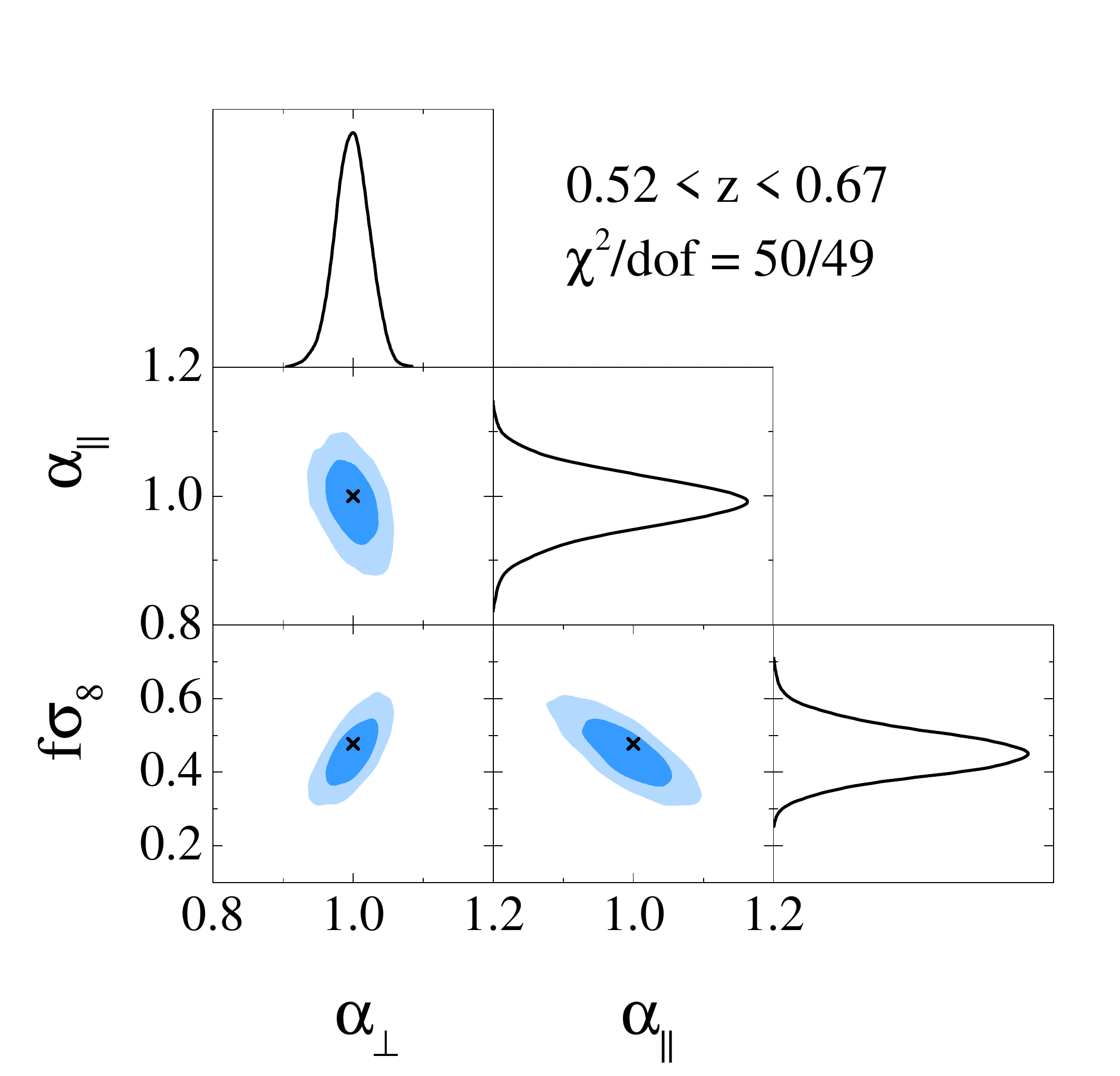}} {\includegraphics[scale=0.22]{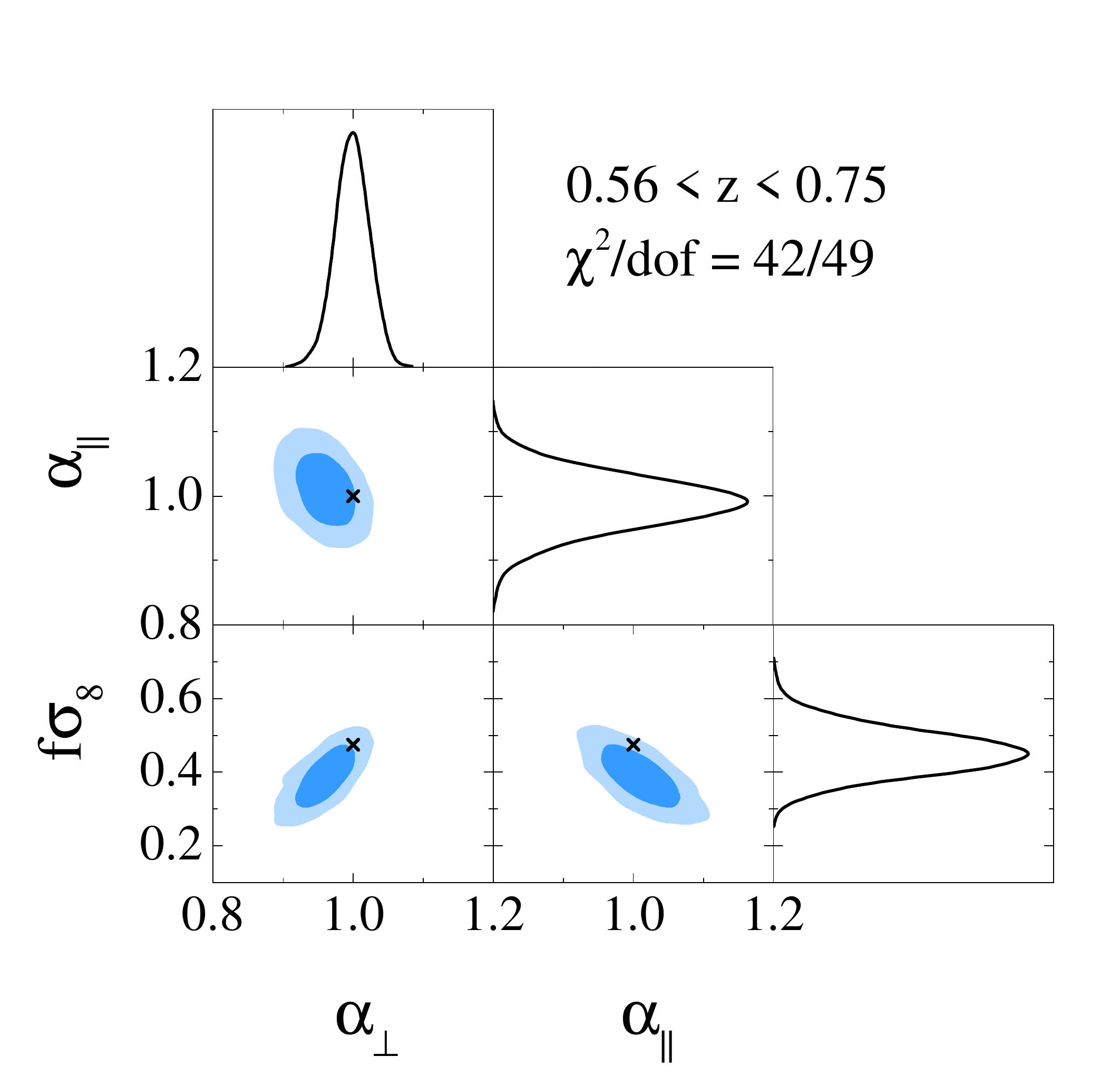}} 
\caption{The measurement in nine redshift bins. We show the one-dimensional posterior distributions and the 68 and 95 \% CL contour plots for parameters $\alpha_{\perp}$, $\alpha_{\parallel}$ and $f\sigma_8$. The black cross in each panel illustrates the fiducial value.}
\label{fig:zbin123-1D-2D}
\end{figure*}

\section{Methodology} 
\label{Method}
\subsection{Theoretical model} 
We use the ``Gaussian streaming model'' (GSM) developed in \citep{Reid2011} to compute the theoretical correlation function. The streaming model has been used to analyze the anisotropic clustering of galaxies measured from the BOSS DR9, DR11 and DR12 samples \citep{Reid2012, Samushia2014, Pellejero2016, Chuang2016, Satpathy2017}.

In the GSM model, the redshift-space correlation function, $\xi^{s}(s_{\perp}, s_{\parallel})$ is given by \citep{Reid2011},
\ba
1+\xi^{s}(s_{\perp}, s_{\parallel}) &=& \int \frac{\d y }{\sqrt{2\pi \left[\sigma^2_{12}(r, \mu)+\sigma^2_{\rm FOG}\right]}}  \left[1+\xi(r)\right] \nonumber \\
&\times& \exp \left\{-\frac{\left[s_{\parallel} - y - \mu v_{12}(r)\right]^2}{2 \left[\sigma^2_{12}(r, \mu)+\sigma^2_{\rm FOG}\right] }\right\} \,,
\label{eq:streaming}
\ea
where $\xi(r)$ is the real-space correlation function computed using the Lagrangian perturbation theory (LPT) \citep{Matsubara2008}. $v_{12}(r)$ is the mean infall velocity of galaxies separated by the real-space distance $r$, and $\sigma_{12} (r, \mu)$ is the pairwise velocity dispersion of galaxies. $v_{12}(r)$ and $\sigma_{12} (r, \mu)$ are calculated using the standard perturbation theory (SPT)  \citep{Bernardeau2002} (see Appendix A in \citep{Reid2011} for details). $y$ is the real-space pair separation along the LOS and $\mu=y/r$. The parameter $\sigma^2_{\rm FOG}$ accounts for the motions of galaxies (see \citet{Reid2012} for details). 

\begin{table}
\caption{The mean value with 68\% CL error of the anisotropic BAO parameters, $\alpha_{\perp}$ and $\alpha_{\parallel}$ and the large-scale RSD parameter, $f \sigma_8$ derived from mock catalogues. The values of $f \sigma_8$ for nine redshift slices in the fiducial cosmology are shown in the last column.}
\begin{center} 
\begin{tabular}{ccccc}
\hline\hline
$z_{\rm eff}$  &$\alpha_{\perp}$ &$\alpha_{\parallel}$  & $f \sigma_8$ & $ \left(f \sigma_8\right)_{\rm fid}$ \\ \hline
$0.31$	&	$1.003	\pm	0.037$	&	$0.988	\pm	0.063$	&	$0.469	\pm	0.098$	&	0.479	\\
$0.36$	&	$0.999	\pm	0.036$	&	$0.991	\pm	0.065$	&	$0.474	\pm	0.097$	&	0.48	\\
$0.40$	&	$0.998	\pm	0.031$	&	$0.998	\pm	0.058$	&	$0.473	\pm	0.086$	&	0.481	\\
$0.44$	&	$1.000	\pm	0.028$	&	$0.993	\pm	0.053$	&	$0.481	\pm	0.076$	&	0.482	\\
$0.48$	&	$1.001	\pm	0.024$	&	$0.989	\pm	0.048$	&	$0.482	\pm	0.067$	&	0.482	\\
$0.52$	&	$1.001	\pm	0.024$	&	$0.986	\pm	0.048$	&	$0.488	\pm	0.065$	&	0.481	\\
$0.56$	&	$1.000	\pm	0.025$	&	$0.988	\pm	0.047$	&	$0.482	\pm	0.067$	&	0.479	\\
$0.59$	&	$1.000	\pm	0.026$	&	$0.990	\pm	0.047$	&	$0.481	\pm	0.066$	&	0.477	\\
$0.64$	&	$1.005	\pm	0.028$	&	$0.982	\pm	0.049$	&	$0.486	\pm	0.070$	&	0.475	\\    \hline          
 \hline                      
\end{tabular}
\end{center}
\label{tab:mock_test}
\end{table}

A fiducial cosmology is assumed to convert the observables of the angular coordinates and redshifts for galaxies into distances. If the fiducial cosmology is different from the true one, it yields geometric distortions parallel and perpendicular to the LOS directions, giving rise to an anisotropy in the galaxy clustering, which is the Alcock-Paczynski (AP) effect \citep{Alcock1979}. With the AP effect, the theoretical correlation function in Eq. (\ref{eq:streaming}) should be revised as,
\begin{equation}
\widehat{\xi}^{s} (s'_{\perp}, s'_{\parallel}) = \xi^{s} (\alpha_{\perp}s_{\perp}, \alpha_{\parallel}s_{\parallel}) \,,
\end{equation}
using two scaling parameters,
\begin{equation}
\alpha_{\perp} =\frac{D_A(z)r_d^{\rm fid}}{D^{\rm fid}_A(z)r_d},\quad\,\alpha_{\parallel}= \frac{H^{\rm fid}(z)r_d^{\rm fid}}{H(z)r_d} \,,
\end{equation}
where, $r_d$ is the sound horizon at the baryon-drag epoch. $D^{\rm fid}_A(z)$ and $H^{\rm fid}(z)$ are the angular diameter distance and Hubble expansion rate in the fiducial cosmology, respectively.

\begin{table*}
\caption{The best-fit value with 68\% CL error for parameters, $\alpha_{\perp}$,  $\alpha_{\parallel}$ and $f \sigma_8$ derived from the BOSS DR12 galaxy catalogue. We also show the derived parameters, \big\{$D_A\times\left(r_d^{\rm fid}/r_d\right), H\times\left(r_d/r_d^{\rm fid}\right)$\big\} or \big\{$D_V\times\left(r_d^{\rm fid}/r_d\right), F_{\mathrm{AP}}$\big\} .}
\begin{center} 
\begin{tabular}{cccccccc}
\hline\hline
$z_{\rm eff}$  &$\alpha_{\perp}$ &$\alpha_{\parallel}$  & $f \sigma_8$  &$D_A\times\left(r_d^{\rm fid}/r_d\right)$ & $H\times\left(r_d/r_d^{\rm fid}\right)$   &  $D_V\times\left(r_d^{\rm fid}/r_d\right)$  & $F_{\rm AP}$  \\ 
 & &   &    & (Mpc) &  ($\rm km\, s^{-1} Mpc^{-1}$)  &  (Mpc)  &    \\ 
$0.31$	&$	0.958	\pm	0.031	$&$	1.029	\pm	0.053	$&$	0.384	\pm	0.083	$&$	930	\pm	29	$&$	77.5	\pm	4.1	$&$	1214	\pm	27	$&$	0.315	\pm	0.022	$\\
$0.36$	&$	0.959	\pm	0.035	$&$	1.012	\pm	0.068	$&$	0.409	\pm	0.098	$&$	1025	\pm	38	$&$	81.2	\pm	5.9	$&$	1376	\pm	40	$&$	0.378	\pm	0.034	$\\
$0.40$	&$	0.972	\pm	0.028	$&$	1.019	\pm	0.057	$&$	0.461	\pm	0.086	$&$	1113	\pm	32	$&$	82.5	\pm	4.9	$&$	1526	\pm	34	$&$	0.430	\pm	0.032	$\\
$0.44$	&$	0.979	\pm	0.021	$&$	1.066	\pm	0.041	$&$	0.426	\pm	0.062	$&$	1190	\pm	26	$&$	80.8	\pm	3.2	$&$	1694	\pm	26	$&$	0.463	\pm	0.024	$\\
$0.48$	&$	1.009	\pm	0.022	$&$	1.024	\pm	0.042	$&$	0.458	\pm	0.063	$&$	1285	\pm	28	$&$	86.2	\pm	3.6	$&$	1827	\pm	28	$&$	0.547	\pm	0.030	$\\
$0.52$	&$	1.011	\pm	0.023	$&$	1.009	\pm	0.055	$&$	0.483	\pm	0.075	$&$	1338	\pm	30	$&$	89.4	\pm	5.4	$&$	1932	\pm	35	$&$	0.607	\pm	0.045	$\\
$0.56$	&$	1.012	\pm	0.022	$&$	0.960	\pm	0.052	$&$	0.472	\pm	0.063	$&$	1386	\pm	30	$&$	96.3	\pm	5.1	$&$	2004	\pm	36	$&$	0.693	\pm	0.045	$\\
$0.59$	&$	0.997	\pm	0.024	$&$	0.999	\pm	0.043	$&$	0.452	\pm	0.061	$&$	1406	\pm	34	$&$	94.3	\pm	4.1	$&$	2113	\pm	35	$&$	0.705	\pm	0.041	$\\
$0.64$	&$	0.968	\pm	0.028	$&$	1.000	\pm	0.037	$&$	0.379	\pm	0.054	$&$	1412	\pm	41	$&$	96.8	\pm	3.5	$&$	2192	\pm	41	$&$	0.746	\pm	0.040	$\\
\hline \hline                      
\end{tabular}
\end{center}
\label{tab:result}
\end{table*}

\begin{figure}
\centering
{\includegraphics[scale=0.3]{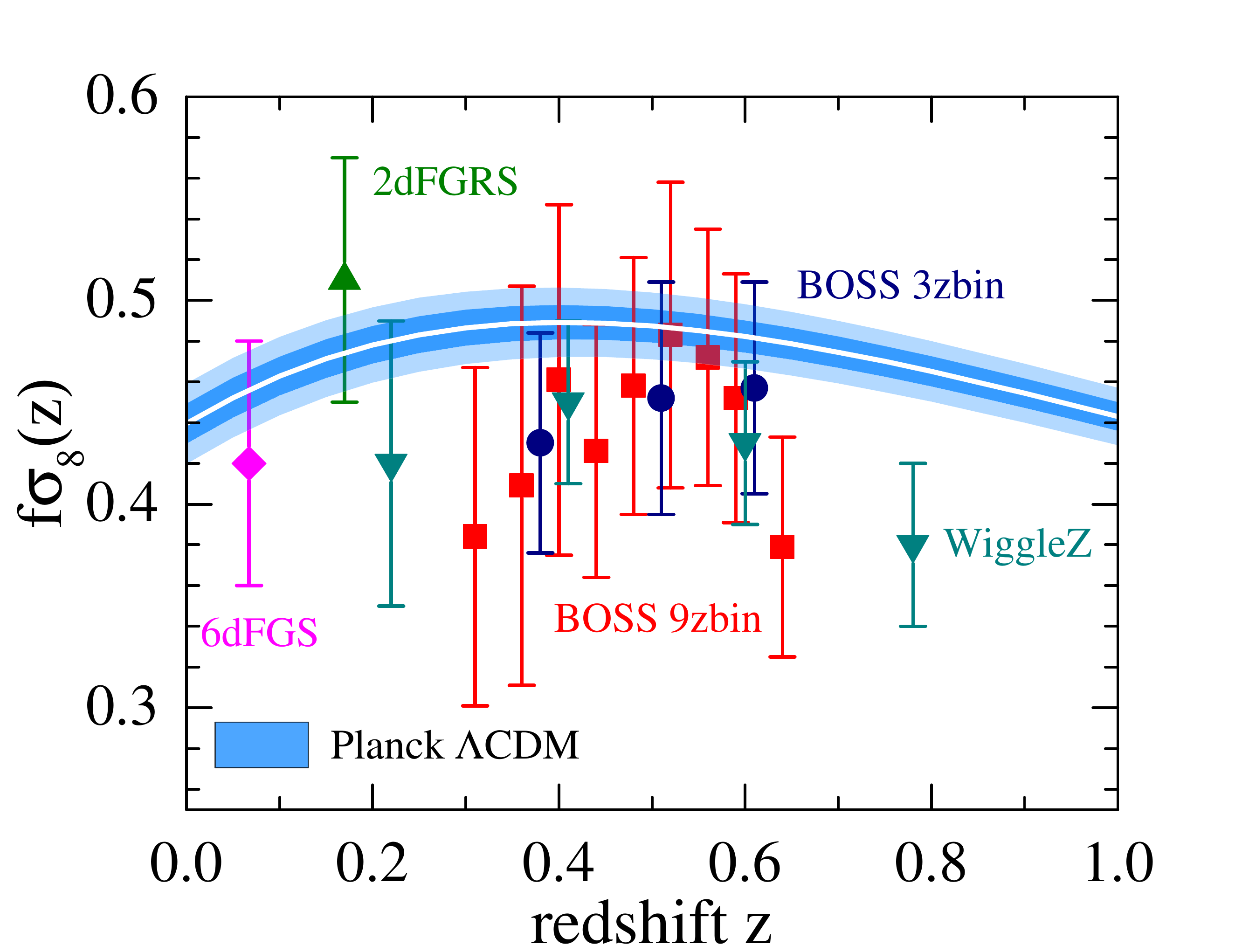}} 
\caption{The $f\sigma_8$ measurement from 2dFGRS \citep{Percival2004}, 6dFGS \citep{Beutler2012}, WiggleZ \citep{Blake2011}, BOSS DR12 3$z$bin measured using correlation function \citep{Satpathy2017} and BOSS DR12 9$z$bin (our tomographic measurements). The light and dark blue shaded bands are the 68 and 95\% CL prediction from Planck assuming the $\Lambda$CDM model \citep{planck}.}
\label{fig:plc-rsd}
\end{figure}

\begin{figure*}
\centering
{\includegraphics[scale=0.2]{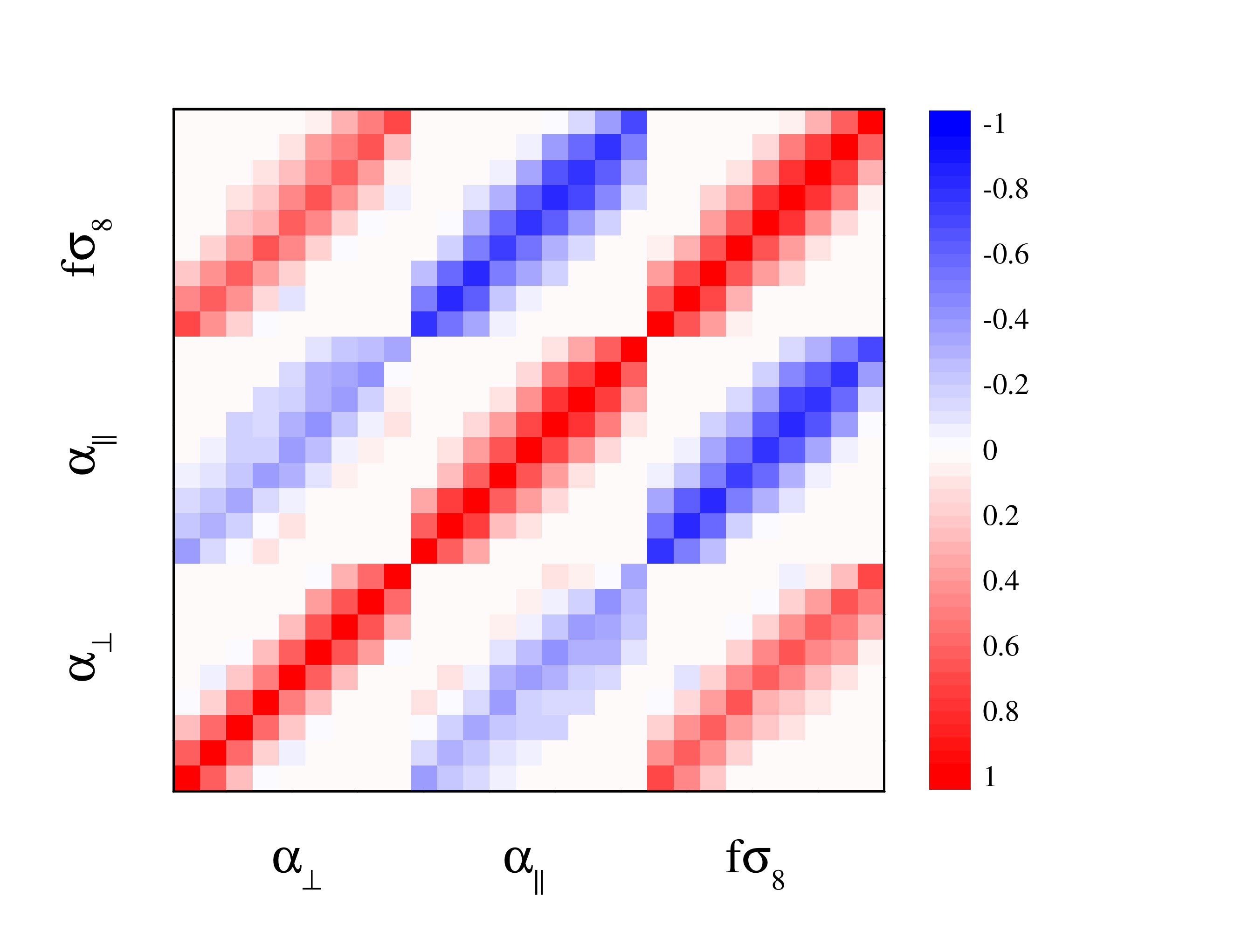}} {\includegraphics[scale=0.2]{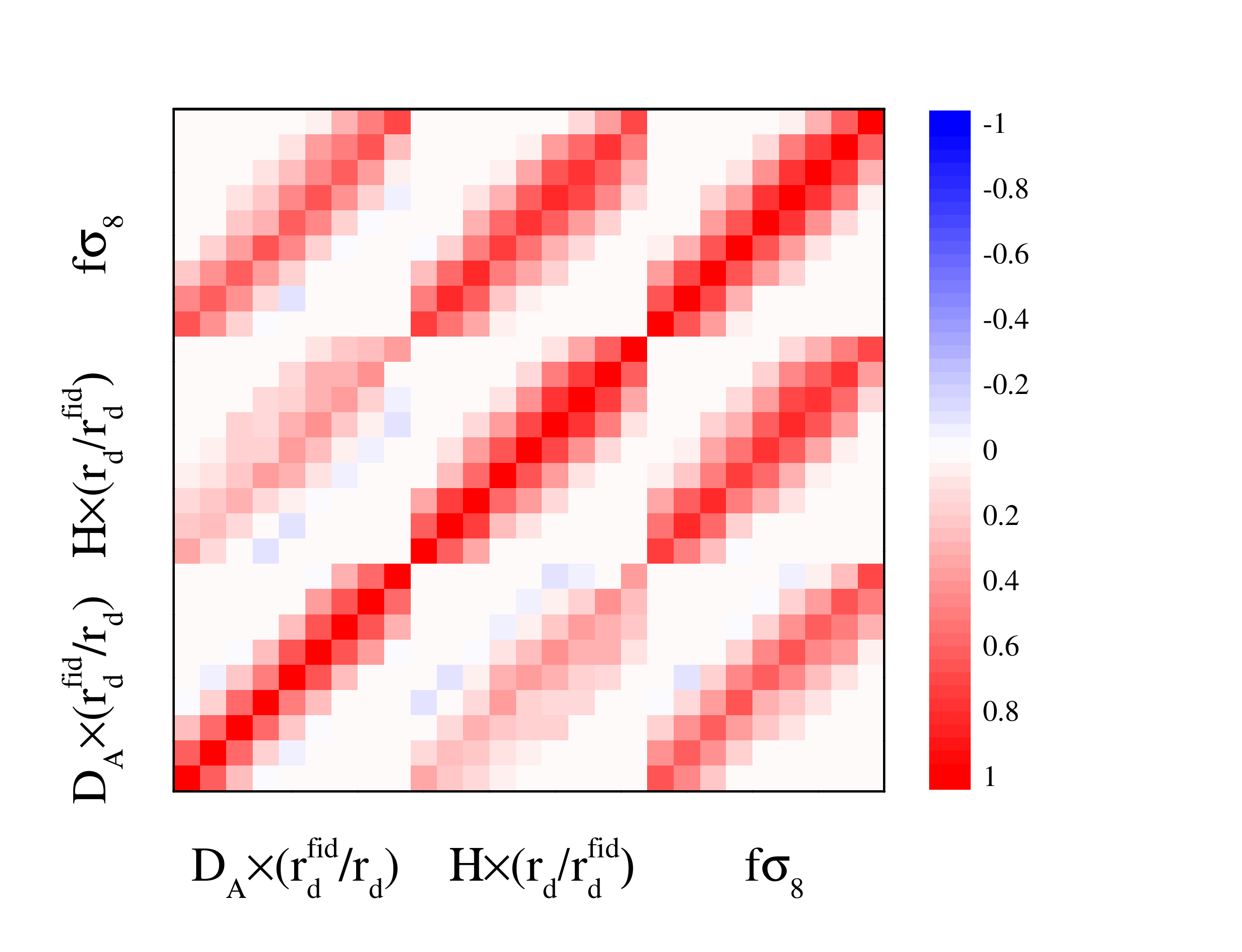}} {\includegraphics[scale=0.2]{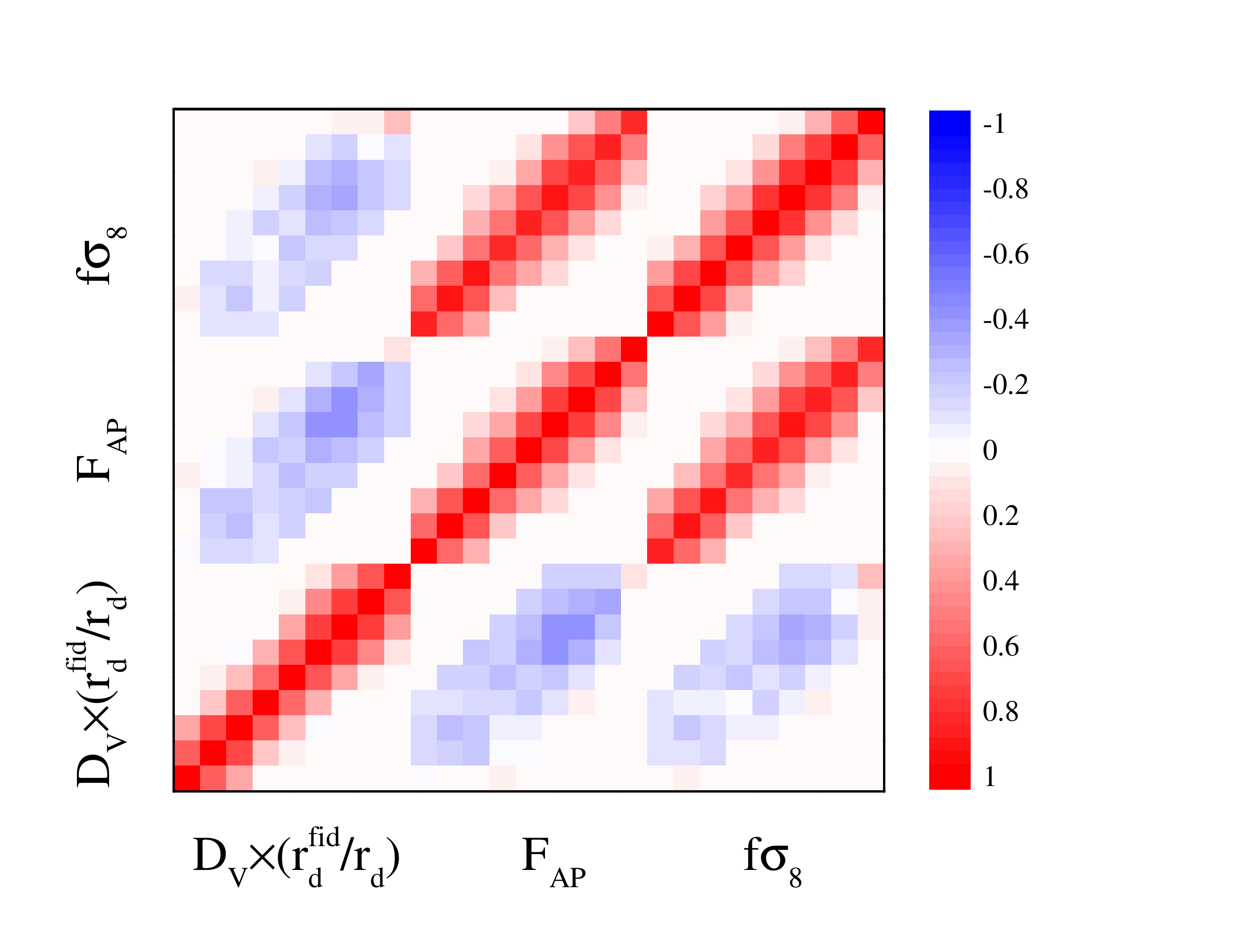}} 
\caption{The correlation coefficient for the measurements, \{$\alpha_{\perp}, \alpha_{\parallel}, f\sigma_8$\} (left panel), \{$D_A\times(r_d^{\rm fid}/r_d), H\times(r_d/r_d^{\rm fid})$\} (middle panel), and\{$D_V\times\left(r_d^{\rm fid}/r_d\right), F_{\mathrm{AP}}$\} (right panel) in nine redshift slices.}
\label{fig:corr_paras}
\end{figure*}

\begin{figure}
\centering
{\includegraphics[scale=0.3]{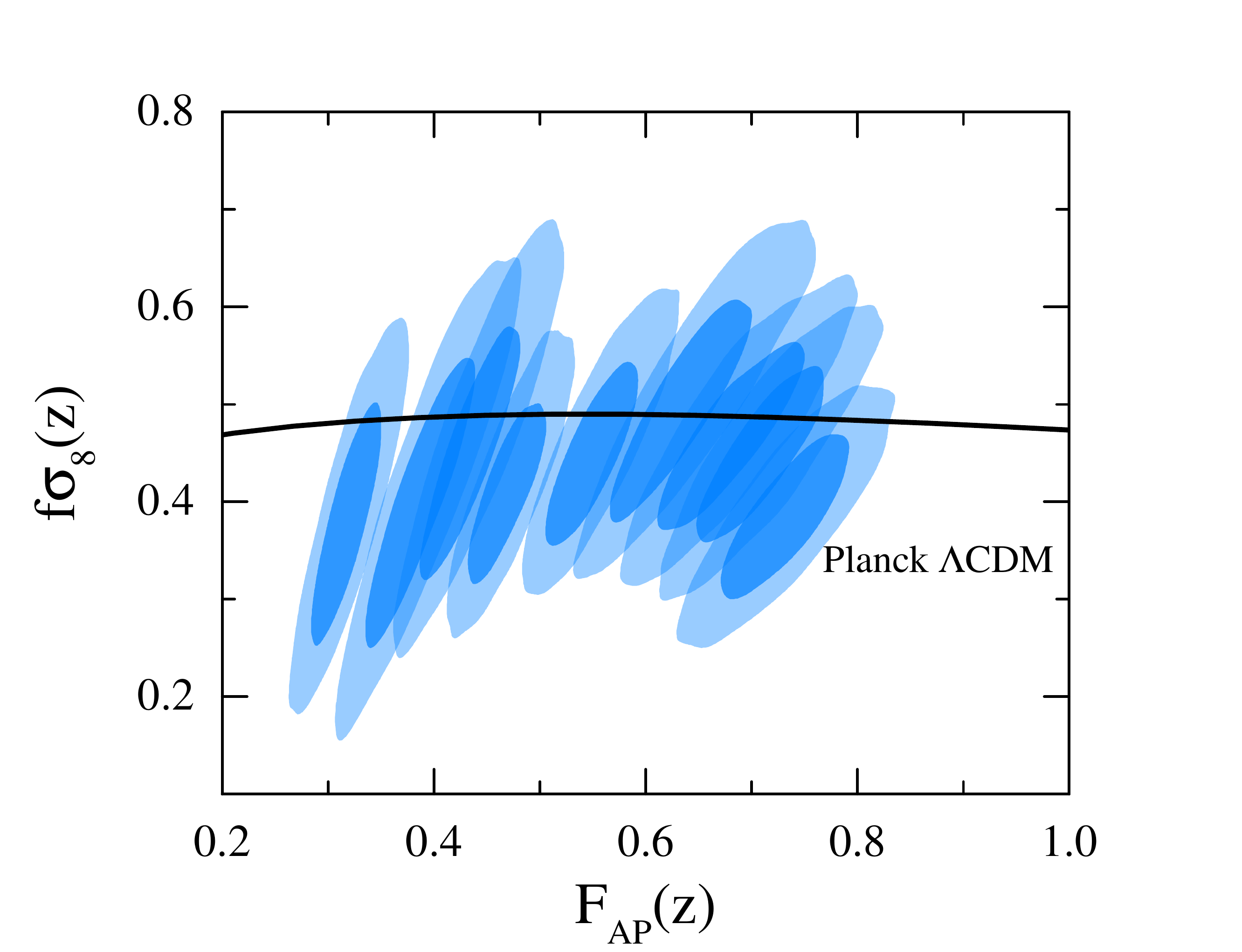}}
\caption{The 68 and 95 \% CL contour plots for parameters, $f\sigma_8$ and $F_{\mathrm{AP}}$ in nine redshift slices. The black solid line shows the best fit model predicted from Planck in the $\Lambda$CDM framework.}
\label{fig:FAP-fs8}
\end{figure}

\subsection{Parameter estimation}
We compute theoretical predictions for the monopole and quadrupole correlation function using the {\tt CosmoXi2D} code \footnote{\url{http://mwhite.berkeley.edu/CosmoXi2D}} \citep{Reid2011}, and use a modified version of {\tt CosmoMC}\footnote{\url{http://cosmologist.info/cosmomc/}} \citep{cosmomc} for parameter estimation. We sample the parameter space of,
\begin{equation}
\bold{p} \equiv \{\alpha_{\perp}, \alpha_{\parallel}, b\sigma_8, f\sigma_8, \sigma^2_{\rm FOG}\} \,,
\end{equation}
with uniform priors of $\alpha_{\perp}\in[0.8, 1.2]$, $\alpha_{\parallel}\in[0.8, 1.2]$, $ b\sigma_8\in[1.0, 2.0]$, $ f\sigma_8\in[0.0, 1.0]$ and $ \sigma^2_{\rm FOG}\in[0.0, 50.0]$ Mpc$^2$, which are conservative priors.

The $\chi^2$ is constructed as follows,
\begin{equation}
\chi^2 (\bold{p}) \equiv  \sum_{i,j}^{\ell,\ell'}  \left[\xi^{\mathrm{th}}_{\ell} (s_i, \bold{p}) -\xi^{\mathrm{obs}}_{\ell}(s_i) \right] F^{\ell,\ell'}_{ij} \left[\xi_{\ell'}^{\mathrm{th}}(s_j, \bold{p}) -\xi^{\mathrm{obs}}_{\ell'}(s_j)\right] \\,
\end{equation}
where $F^{\ell,\ell'}_{ij}$ is the inverse of the data covariance matrix estimated for 2045 mocks (see \citet{Wang2017} for details). The correlation functions are measured with a bin width of 5 $\mpcoh$ on scales of $25 -160 \mpcoh$. 

\section{Results} 
\label{result}
\subsection{Mock tests}
\label{sec:mock}
We perform tests on the MD-Patchy mock catalogues by fitting the average of 2045 mocks. The result is shown in Table \ref{tab:mock_test}, where we present the mean value with 68\% confidence level (CL) uncertainties of parameters $\alpha_{\perp}$, $\alpha_{\parallel}$ and $f \sigma_8$. The fiducial cosmology we use here corresponds to the input cosmology of the mocks, therefore we expect that the average values of parameters $\alpha_{\perp}$ and $\alpha_{\parallel}$ are equal to $1$. The values of $f \sigma_8$ in the fiducial cosmology are shown in the last column of Table \ref{tab:mock_test}. In left panels of Figure \ref{fig:combias}, we show the difference (with 68\% CL uncertainty) between the mean and the expected values of parameters $\alpha_{\perp}$, $\alpha_{\parallel}$ and $f \sigma_8$ respectively, while the panels on the right show the significance of the bias in terms of the 68\% CL uncertainty. For $\alpha_{\perp}$, the mean values are in good agreement with unity, and there is a shift of $0.005$ towards higher values in the worst case (for the last redshift bin). It is substantially smaller than the statistical uncertainty, which is $0.028$. The largest bias in terms of the uncertainty is $0.18\, \sigma$ in the $\alpha_{\perp}$ parameter, as shown at the top right-hand panel of Figure \ref{fig:combias}. Taking this shift into account by adding the bias to the statistical error in quadrature, we find that the total uncertainty of $\alpha_{\perp}$ gets increased by $1.6$\%. The largest shift on the $\alpha_{||}$ parameter is $0.018$ towards lower values also in the last redshift bin, which is $0.37\, \sigma$ (the middle right-hand panel of Figure \ref{fig:combias}). This shift would increase the total uncertainty by $6.5$\%. The value of the $f \sigma_8$ parameter is shifted in the worst case towards higher values of $0.011$, which represents $0.16\, \sigma$ as shown at the bottom right-hand panel of Figure \ref{fig:combias}. The shift slightly increase the total uncertainty by $1.2$\%. Overall, we can reproduce the input parameters in the fiducial cosmology within a shift of 0.5\% for $\alpha_{\perp}$, 1.8\% for $\alpha_{||}$, and 1.1\% for $f \sigma_8$ respectively.

\subsection{Measurements from the data catalogue}
\label{sec:data}
We present the measurement from the DR12 catalogue in Table \ref{tab:result}, showing the best-fit value with 68\% CL uncertainties for parameters, $\alpha_{\perp}$,  $\alpha_{\parallel}$ and $f \sigma_8$ in each redshift slice. The one-dimensional posterior distributions and the 68 and 95\% CL contour plots for these three parameters are shown in Figure \ref{fig:zbin123-1D-2D}. In the two-dimensional contour plots, we show the fiducial values of the parameters in black crosses. 

Figure \ref{fig:plc-rsd} presents our measurements of $f\sigma_8$ at different redshifts together with various other measurements, including Planck \citep{planck}, 2dFGRS \citep{Percival2004}, 6dFGS \citep{Beutler2012}, BOSS \citep{Satpathy2017} and WiggleZ \citep{Blake2011}. To compare our measurement from the RSD measurement using the same galaxy sample in three redshift slices, as presented in \cite{Satpathy2017}, we compress our measurements on $f\sigma_8$ into three redshift bins. We compress the first 4 redshift bins covering $z\in[0.2,0.51]$ into one measurement by performing a fit as below,
\begin{equation}
\chi^2=\left(\overline{\theta}-\theta_i \right) C^{-1}\left(\overline{\theta}-\theta_i\right)^T\,,
\end{equation}
where, $\overline{\theta}$ is a single parameter within the redshift range of 0.2<z<0.51. $\theta_i$ denote the measurements of $f\sigma_8$ in the first 4 redshift bins. $C$ is the covariance matrix between the $f\sigma_8$ measurements in the first 4 redshift bins with other parameters marginalized over. In the same way, the measurements in the 5th and 6th $z$bins are compressed into one measurement and the remaining $z$ bins are compressed into the last measurement. We find that $f\sigma_8=0.403 \pm 0.049~(0.2<z<0.51);~f\sigma_8=0.462 \pm 0.063~(0.4<z<0.59)$ and $f\sigma_8=0.413 \pm 0.047~(0.48<z<0.75)$. These results are in agreement with that in \citet{Satpathy2017}, where the measurements of correlation function multipoles were used and the GSM in theoretical framework of the Convolution Lagrangian Perturbation Theory (CLPT) \citep{Carlson2013, Wang2014} was adopted. 

As our redshift slices significantly overlap with each other, the errors of our measurements from various redshift slices are expected to correlate as well. To quantify the correlation, we perform a joint fit on parameters for all pairs of overlapping redshift bins simultaneously following \citet{Zhao2017}, and calculate the correlation matrix of parameters, $\{\alpha_{\perp}, \alpha_{\parallel}, f \sigma_8\}$ in nine redshift slices. The result is shown in the left panel of Figure \ref{fig:corr_paras}. We can see that the auto-correlation of the same parameter and the cross-correlation between different parameters decrease as the redshift separation increases. The parameter $\alpha_{\perp}$ anti-correlates with $\alpha_{\parallel}$, and positively correlates with $f \sigma_8$. The parameters $\alpha_{\parallel}$ and $f \sigma_8$ negatively correlate, which is as expected.

Given $\alpha_{\perp}$ and $ \alpha_{\parallel}$, we derive the angular diameter distance, $D_A(z) r_d^{\rm fid}/r_d$ and Hubble parameter, $H(z) r_d/ r_d^{\rm fid}$, shown in Table \ref{tab:result}. The correlation matrix of parameters, $\{D_A r_d^{\rm fid}/r_d, Hr_d/ r_d^{\rm fid}, f \sigma_8\}$ in nine redshift slices is presented in the middle panel of Figure \ref{fig:corr_paras}. It is seen that they positively correlate with each other, which is expected.

We present the result in another parametrization, given by $\{D_V r_d^{\rm fid}/r_d, F_{\mathrm{AP}}, f \sigma_8\}$ where the effective volume distance $D_V(z)\equiv[cz(1+z)^2D_A^2(z)H^{-1}(z)]^{1/3}$ and the AP parameter $F_{\mathrm{AP}}\equiv(1+z)D_A(z)H(z)/c$, in Table \ref{tab:result}. Their correlation matrix in nine redshift slices is shown in the right panel of Figure \ref{fig:corr_paras}. There is a clear positive correlation between the AP parameter and $f \sigma_8$ while the parameters $D_V r_d^{\rm fid}/r_d$ and $f \sigma_8$ are nearly uncorrelated. In the two-dimensional contour plots for parameters $f\sigma_8$ and $F_{\mathrm{AP}}$, shown in Figure \ref{fig:FAP-fs8}, our measurement is consistent with the Planck $\Lambda$CDM prediction within the 95\% CL region for all redshift slices.

\begin{figure}
\centering
{\includegraphics[scale=0.3]{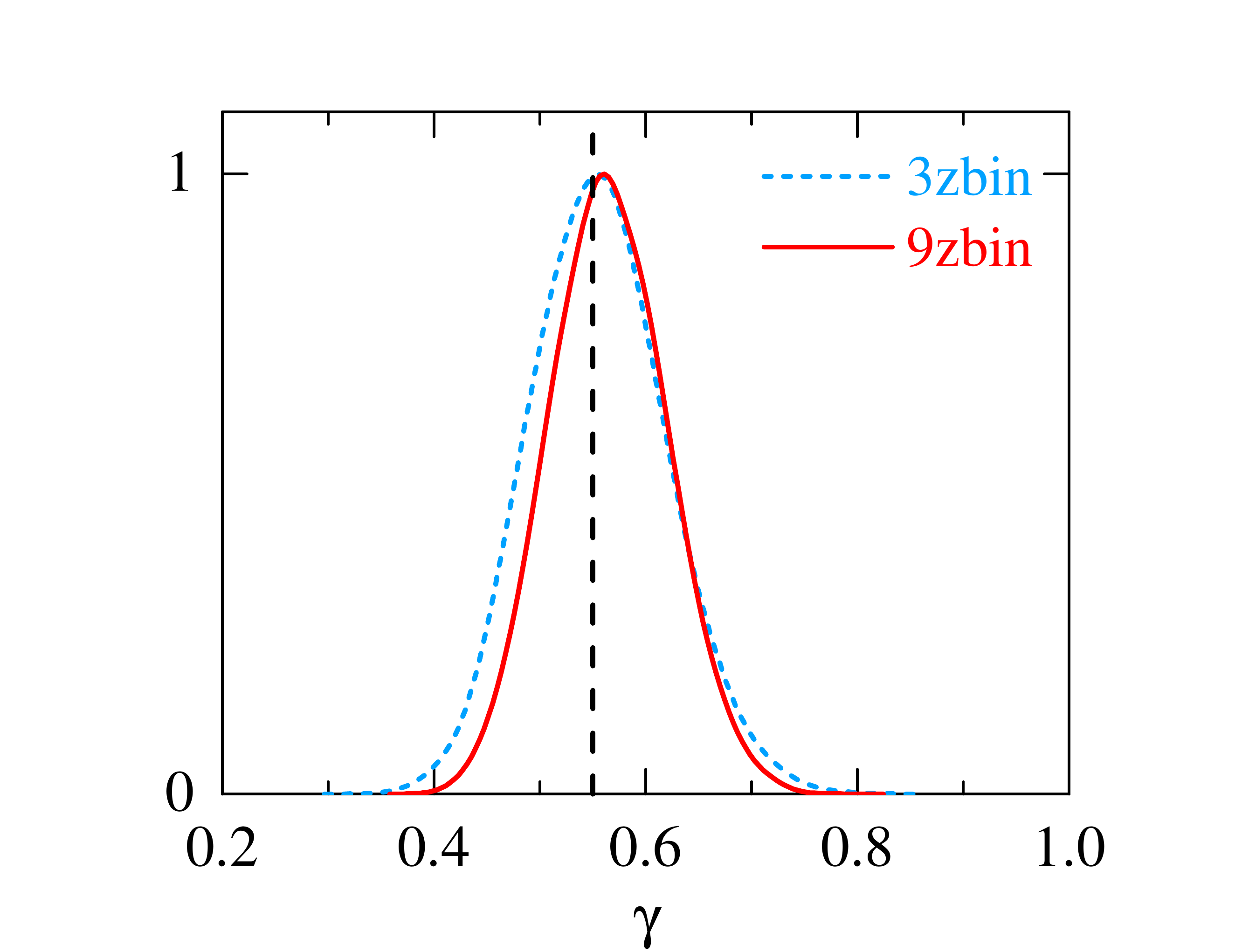}}
\caption{The one-dimensional posterior distribution of $\gamma$ derived from the mock data of ``$3~z {\rm bins}$'' (blue short-dashed) and ``$9~z {\rm bins}$'' (red solid) respectively (see text for details). The vertical black dashed line illustrates the $\Lambda$CDM prediction as a reference.}
\label{fig:gamma}
\end{figure}

\begin{figure}
\centering
{\includegraphics[scale=0.3]{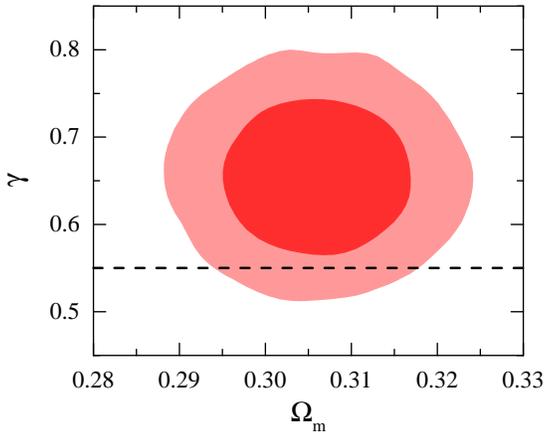}}
\caption{The 68 and 95 \% CL contour plot for parameters $\gamma$ and $\Omega_m$ using a combined dataset of Planck + SNe + BAO + RSD (``9 $z$bin"). The horizontal black dashed line illustrates the $\Lambda$CDM prediction as a reference.}
\label{fig:gamma_om}
\end{figure}

\subsection{Constraints on Modified Gravity} 
Using our tomographic BAO and RSD measurements  \footnote{The measurements and covariance matrices are available at \url{https://github.com/ytcosmo/TomoBAORSD}}, we perform an observational constraint on a phenomenological model of Modified Gravity (MG), which is parametrised by the gravitational growth index, $\gamma$, \citep{Linder2005, Linder2007} as, 
\begin{equation} \label{eq:growth_rate}
f(a)=\Omega_m(a)^{\gamma} \,, 
\end{equation}
where $f(a)$ is the growth rate function of the scale factor $a$, and $\Omega_m(a)$ is the dimensionless matter density, \ie, $\Omega_m(a) = 8 \pi G \rho_m(a)/ 3 H^2(a)$. In the framework of General Relativity (GR), the value of $\gamma$ is expected to be $6/11$ \citep{Linder2005}. Given $f(a)$, the growth factor $D$ can be calculated via,
\begin{equation} \label{eq:growth_factor}
D(a)=\mathrm{exp} \left[- \int_a^1 \d a' f(a') /a' \right]  \,, 
\end{equation}
which is used to compute $\sigma_8(z)$ through
\begin{equation} \label{eq:sig8}  
\sigma_8(z) =\sigma_8(z=0) \frac{D(z)}{D_{\mathrm{GR}}(z)} \frac{D_{\mathrm{GR}}(z_{\mathrm{ini}})}{D(z_{\mathrm{ini}})} \,.
\end{equation}
Here we assume an initial redshift $z_{\mathrm{ini}} = 50$, where the modification of gravity starts to take effect.

Given Eqs. (\ref{eq:growth_rate}) - (\ref{eq:sig8}), we can compute theoretical predictions of $f\sigma_8 (z)$ at any given redshifts. We then perform an estimation of cosmological parameters with a modified version of {\tt CosmoMC} \citep{cosmomc}, sampling the following parameter space,
\begin{equation}
\bold{P} \equiv \{\omega_{b}, \omega_{c}, \Theta_s, \tau, n_s, A_s, \gamma \} \,,
\end{equation}
where $\omega_{b}$ and $\omega_{c}$ are the densities of baryon and cold dark matter, $\Theta_s$ is the ratio of the sound horizon to the angular diameter distance at the decoupling epoch (multiplied by $100$), $\tau$ is the optical depth, $n_s$ and $A_s$ are the spectral index and the amplitude of the primordial power spectrum, and $\gamma$ is the growth index. We fix the sum of the neutrino mass to $0.06$ eV and assume an effective number of relativistic species, $N_{\rm eff} = 3.046$.

We use a combined data set, including the temperature and polarisation power spectra from Planck 2015 data release \citep{planck}, the ``Joint Light-curve Analysis'' sample of type Ia SNe \citep{SNeJLA}, and a joint measurement of BAO and RSD from the BOSS DR12 completed sample in nine tomographic slices reported in this work (``9 $z$bins''). 

Before measuring cosmological parameters including $\gamma$ using the actual DR12 BAO and RSD measurement presented in Sec. \ref{sec:data}, we perform a consistent test to validate our pipeline using the BAO and RSD measurement derived from the DR12 mock catalogue (see Sec. \ref{sec:mock} for details of the mock catalogue). With a combined dataset of Planck + SNe + BAO (mock) + RSD (mock) \footnote{Note that the cosmology used to generate the mock galaxy catalogue is derived from the Planck data, which is also consistent with the SNe data in terms of the background cosmological parameters. We removed the ISW part of the Planck data in this calculation so that neither the Planck nor SNe data are informative to infer $\gamma$.}, we find that $\gamma=0.565 \pm 0.054$. To quantify the information gain from the tomographic BAO and RSD analysis, we compress our nine-bin measurements into those at three redshift bins, and we denote the compressed measurement as ``3 $z$bins''. We then repeat the mock test using ``3 $z$bins'' instead and find that $\gamma=0.556\pm 0.063$. Figure \ref{fig:gamma} compares the one-dimensional posterior distribution of $\gamma$ derived from the mock data in both cases (with Planck and SNe datasets combined). As shown, both results agree well with the $\Lambda$CDM prediction of $\gamma=0.545$ within the uncertainty, which validates our analysis, and the constraint on the precision of $\gamma$ gets improved by $14\%$. This is because ``9 $z$bins'' is more informative in redshift, which helps with the constraint on $\gamma$ as a variation of $\gamma$ changes $f\sigma_8$ at different redshifts in different ways \footnote{The constraint on $\Omega_m$ also gets improved, but less significantly, namely, $\Omega_m=0.3139\pm0.0071$ (3 $z$bins) and $\Omega_m=0.3148\pm0.0070$ (9 $z$bins). The Figure-of-Merit (FoM) between $\Omega_m$ and $\gamma$, which is inversely proportional to the area of the 68\% CL contour, is improved by 18\%.} .

\begin{figure}
\centering
{\includegraphics[scale=0.3]{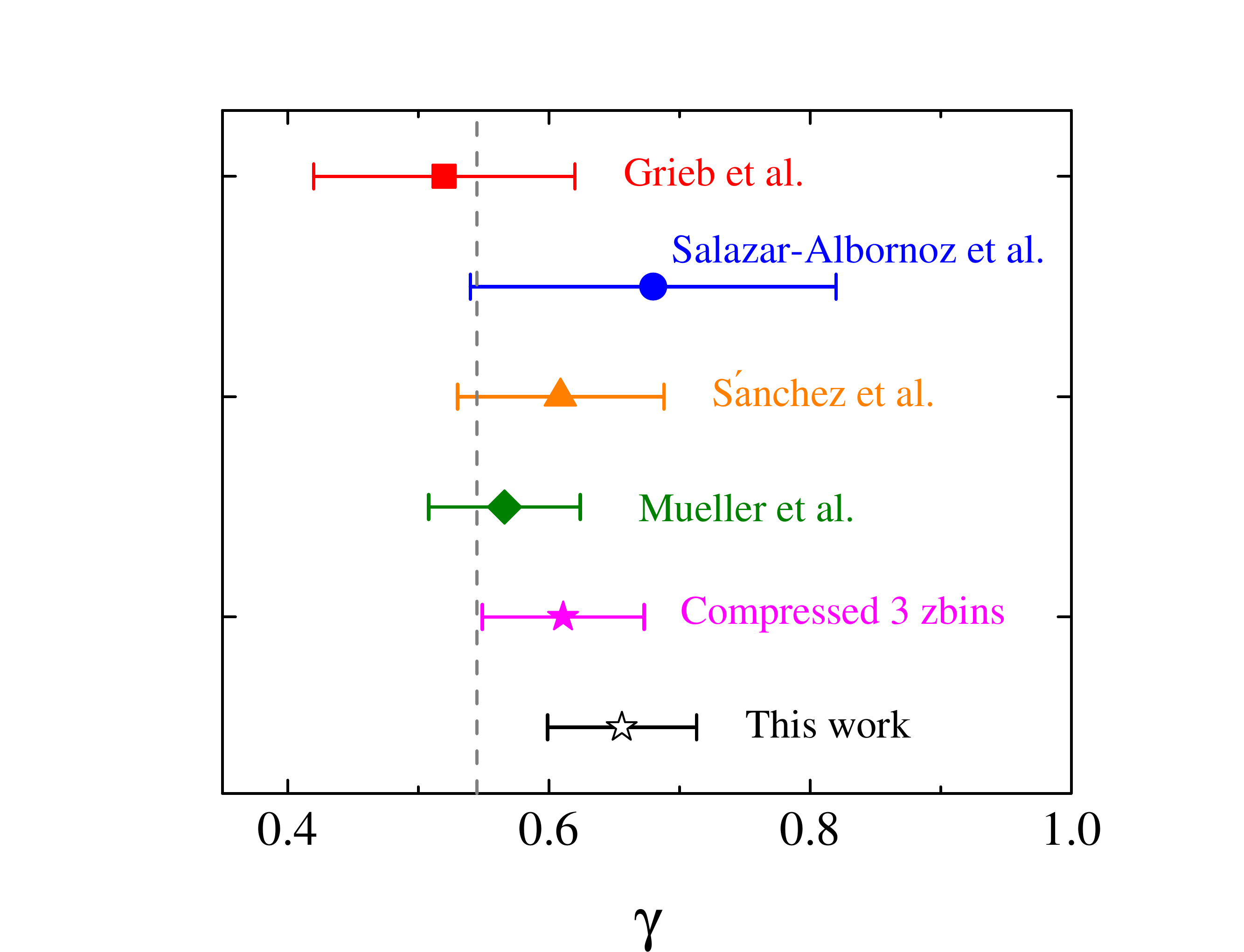}}
\caption{The constraint on the $\gamma$ parameter using our measurement in comparison with those in other BOSS DR12 papers \citep{Grieb2017, Salazar-Albornoz:2016psd, Ariel2017I, Mueller2016}.}
\label{fig:gamma_BOSS}
\end{figure}

We then apply our analysis to the actual measurement presented in Sec. \ref{sec:data}, and find that $\gamma=0.656 \pm 0.057$, which is consistent with the $\Lambda$CDM prediction within 95\% CL. We show the two-dimensional contour between $\Omega_m$ and $\gamma$ in Figure \ref{fig:gamma_om}. Using the ``3 $z$bins'' measurement compressed from the actual tomographic measurement, we find $\gamma=0.611\pm0.062$, which is slightly looser than that derived from our tomographic RSD measurement. The level of improvement is consistent with that using the mocks. We summarize the constraint on the $\gamma$ parameter derived from other BOSS DR12 papers \citep{Grieb2017, Salazar-Albornoz:2016psd, Ariel2017I, Mueller2016} in Figure \ref{fig:gamma_BOSS}. As shown, our measurement is in excellent agreement with these results within 68\% CL, with a marginal improvement in the uncertainty.

\section{Conclusions} 
\label{conclusion}
We analyse the anisotropic clustering of the BOSS DR12 galaxies and simultaneously constrain the cosmic expansion rate and large-scale structure growth in nine overlapping redshift slices. We work with the galaxy correlation function multipoles and adopt the GSM model to calculate the theoretical predictions. The analysis pipeline is validated using the MD-Patchy mock catalogues, before applied to the DR12 galaxy catalogue. We present the combined measurement of the effective volume distance, $D_V r_d^{\rm fid}/r_d$, the AP parameter, $F_{\mathrm{AP}}$ and the parameter of linear structure growth, $f \sigma_8$. We obtain a precision of $1.5\%-2.9\%$ for $D_V r_d^{\rm fid}/r_d$, $5.2\%-9\%$ for $F_{\mathrm{AP}}$ and $13.3\%-24\%$ for $f \sigma_8$, depending on effective redshifts of the redshift slices. Our measurement on $f \sigma_8$ agrees with the Planck $\Lambda$CDM result within the 95\% CL.

We perform a cosmological implication of our measurement (combined with Planck and SNe) to constrain $\gamma$, the gravitational growth index. We firstly validate our pipeline by reproducing the value of $\gamma$ in the $\Lambda$CDM model by fitting to the BAO and RSD measurement derived from mock data. This mock test also confirms that our tomographic measurement is more informative, for the constraint on $\gamma$, than that with less redshift slices, namely, the uncertainty on $\gamma$ gets tightened by 14\% when ``9 $z$bins'' is used rather than ``3 $z$bins''. We then constrain $\gamma$ using our BAO and RSD measurement from the actual DR12 survey, and find that $\gamma=0.656 \pm 0.057$, which agrees with the $\Lambda$CDM prediction within the 95\% CL.

Admittedly, the information gain from the tomographic RSD measurement of BOSS DR12 is not significant, which is expected for a galaxy survey covering a moderate redshift range. However, there is much richer information on the lightcone from deeper surveys, \eg, DESI \footnote{\url{http://desi.lbl.gov/}} and Euclid \footnote{\url{https://www.euclid-ec.org/}}, which can be extracted using our method to tighten cosmological constraints.
       
\section*{Acknowledgements}
YW is supported by the NSFC Grant No. 11403034, and by the Young Researcher Grant of National Astronomical Observatories, Chinese Academy of Sciences. GBZ is supported by NSFC Grant No. 11673025, and by a Royal Society-Newton Advanced Fellowship. GBZ and YW are supported by National Astronomical Observatories, Chinese Academy of Sciences, and by University of Portsmouth. 

Funding for SDSS-III has been provided by the Alfred P. Sloan Foundation, the Participating Institutions, the National Science Foundation, and the US Department of Energy Office of Science. The SDSS-III web site is \url{http://www.sdss3.org/}. SDSS-III is managed by the Astrophysical Research Consortium for the Participating Institutions of the SDSS-III Collaboration including the University of Arizona, the Brazilian Participation Group, Brookhaven National Laboratory, Carnegie Mellon University, University of Florida, the French Participation Group, the German Participation Group, Harvard University, the Instituto de Astrofisica de Canarias, the Michigan State/Notre Dame/JINA Participation Group, Johns Hopkins University, Lawrence Berkeley National Laboratory, Max Planck Institute for Astrophysics, Max Planck Institute for Extraterrestrial Physics, New Mexico State University, New York University, Ohio State University, Pennsylvania State University, University of Portsmouth, Princeton University, the Spanish Participation Group, University of Tokyo, University of Utah, Vanderbilt University, University of Virginia, University of Washington, and Yale University.

This research used resources of the National Energy Research Scientific Computing Center, which is supported by the Office of Science of the U.S. Department of Energy under Contract No. DE-AC02-05CH11231, the SCIAMA cluster supported by University of Portsmouth, and the ZEN cluster supported by NAOC.

\bibliographystyle{mn2e}
\bibliography{tomoBAORSD}

\label{lastpage}

\end{document}